\documentclass[onecolumn,authoryear]{els-mrw} 

\usepackage{amsmath,amssymb,amsfonts,amsthm,makeidx,graphicx}
\usepackage{txfonts}
\usepackage{helvet}

\usepackage{multirow}
\usepackage[normalem]{ulem}
\useunder{\uline}{\ul}{}

\usepackage{ mathrsfs }

\newcommand{\be}{\begin{equation}}
\newcommand{\ee}{\end{equation}}

\newcommand{\apj}{ApJ}
\newcommand{\apjl}{ApJL}

\newcommand{\aj}{AJ}
\newcommand{\apjs}{ApJS}
\newcommand{\mnras}{MNRAS}
\newcommand{\araa}{ARA\&A}
\newcommand{\aap}{A\&A}
\newcommand{\apss}{Astrophysics and Space Science}
\newcommand{\pasa}{PASA}
\newcommand{\pasp}{PASP}
\newcommand{\nat}{{\it Nature}}

\begin{document}

\chapter{Morphological Classification of Galaxies}\label{chap1}

\author[1]{Karen L. Masters}%
%\author[2]{Second Author}%
%\author[1,2]{Third Author}%

\address[1]{\orgname{Haverford College}, \orgdiv{Departments of Astronomy and Physics}, \orgaddress{370 Lancaster Avenue, Haverford, PA 19041, USA}}
%\address[2]{\orgname{Name of Institute}, \orgdiv{Division or Department}, \orgaddress{Address of Institute}}

%\articletag{Chapter Article tagline: update of previous edition,, reprint..}

\maketitle

%The chapter should be around 10,000 words including references and tables.

\begin{glossary}[Glossary]
\term{Early-type galaxy} a type of galaxy, typically not actively forming stars. Often used synonymously with elliptical and/or lenticular, this type sometimes also includes the earlier spiral types. 

\term{Edge-on or Spindle galaxy} a disc shaped galaxy viewed in the edge-on orientation (inclination of 90$^\circ$). 

\term{Elliptical galaxy} a type of galaxy which is elliptical (or oval) in shape, and otherwise shows no distinct features.

\term{Galactic Bar} a linear structure of stars stretching across the central regions of a galaxy which reveals the presence of a rotating disc.

\term{Galactic Bulge} the central light concentration found in many spiral galaxies. 

\term{Galactic Ring} a ring shapes feature in a galaxy, which may be further subdivided into ``nuclear", ``inner" or ``outer" rings.

\term{Late-type galaxy} a type of galaxy, typically actively forming stars. Often used synonymously with spiral galaxy, or to sub-divide spirals into ``later" or ``earlier" types. 

\term{Spiral arms} spiral shaped features, which may be continuous over large radial ranges, or patchy in nature.

\term{Spiral galaxy} a type of galaxy identified as having spiral shaped features (or spiral arms).

\end{glossary}

\begin{glossary}[Nomenclature]
\begin{tabular}{@{}lp{34pc}@{}}
$B/T$ & Bulge-to-total radius, often used to identify different types of spiral/disc galaxies\\
CAS & Concentration, Assymmetry, clumpinesS - a system of non-parametric measurements on galaxy images which has been used for automated morphological classification\\
{\tt frac\_deV} & The de Vaucouleurs Fraction - a parameter in SDSS photometric pipelines describing how well the light profile of a galaxy matches that of the de Vaucouleurs profile which is best used in elliptical galaxies \\
GZ & Galaxy Zoo\\
$i$ & Inclination, or the angle made between the normal to the disc of a galaxy and the line-of-sight. \\
$n_s$ & Sersic index; a parameter describing the shape of the aximuthally averaged surface brightness profile ($n=1$ is an exponential decline characteristic of disc; $n=1/4$, or ``de Vaucouleur" profile reveals a more concentrated/spheroidal collection of stars\\
$\phi$ & Pitch angle, or the angle of a spiral arm to the tangent of a circle at the same radius \\
RC3 & The Third Reference Catalog of Galaxies\\
$r_e$ & Effective radius - radius enclosing half the light of a galaxy \\
SDSS & The Sloan Digital Sky Survey\\
\end{tabular}
\end{glossary}

\begin{abstract}[Abstract]
The morphological classification of galaxies provides vital physical information about the orbital motions of stars in galaxies, and correlates in interesting ways with star formation history, and other physical properties. Galaxy morphological classification is a field with a history of more than 100 years of development, and many scientists have introduced new classification schemes, resulting in a sometimes confusing array of terminologies and overlapping classes. In this article I provide a brief historical review of galaxy classification, but focus mostly on providing a summary of how the morphological variety of galaxies seen in our expanding Universe are described. I review traditional visual classification, morphometric measurements, crowd-sourcing for large scale visual classifications (Galaxy Zoo), and of course the recent explosion of interest in making use of machine learning techniques for galaxy morphology classification. A look up table is provided for cross matching of various terminologies currently in use for galaxy morphology classification as well as brief definitions of the main morphological types.  
\end{abstract}

\begin{BoxTypeA}[chap1:box1]{Key Points}
\section*{Morphology of Galaxies}
The word ``morphology" is used to describe the study of shape and form of any object. The morphological classification of galaxies is thus classification based on the shapes and structures seen in images of galaxies. 

\section*{Morphology and the Physical Properties of Galaxies}
The physical properties of galaxies (e.g. mass, gas content, star formation rates, angular momentum) vary systematically between different morphological types, and various morphological features can be used to reveal a frozen snapshot of dynamical processes in the orbits of stars (and where present gas) in a galaxy. 

\end{BoxTypeA}

\section{Introduction}\label{sec:intro}
The galaxies we see when we look out into the Universe come in a wide variety of sizes, colours and shapes. This diversity fascinates us with its beauty, but also teaches us something fundamental about the development of galaxies over cosmic time, and the different components that build a galaxy. When faced with a diversity of similar objects, one of the most basic tools of the scientist is classification. There are many different ways to attempt to classify galaxies by their morphology, but galaxy classification started with traditional visual inspection by small numbers (or individual) astronomers. This has developed today into a highly successfully method which use crowd-sourced visual inspection (i.e. Galaxy Zoo\footnote{{\tt www.galaxyzoo.org}}) to provide quantitative visual morphologies. In the era of computing, the use of automated, or semi-automated measurement of either structural parameters, or non-parametric structural measurements, which have been shown to correlate with visual morphology is also common, and in recent years there has been an enormous growth in the use of machine learning techniques to classify galaxies. 

Galaxy morphology at its most basic level, tells us about the physics of assembly of a galaxy. Since the light from galaxies (mostly) comes from the stars, morphology is a frozen snapshot of where those star are, and are forming today in a galaxy. You can also consider galaxy morphology as an (albeit imperfect) proxy for measurements of the orbital motions of the stars - visible discs in a galaxy reveal places where stars are coherently rotating, dynamical features like spirals and bars give even more information on those orbits, while smoother spheroidal blobs reveal the presence of more random motions. 

The intent of this article is to introduce and compare galaxy morphology terminology in common use in the astronomical literature today, with some comments and references as to the physical properties of different types of galaxies. It's intended as a quick reference/guide for students new to the study of galaxies, and to consider how samples of galaxies selected with different methods (e.g. via star formation properties, colour, morphometric measurements, Galaxy Zoo classifications, or traditional classification types) might compare to each other.

\section{Classification of Galaxies}\label{sec:intro}

\begin{figure}[b]
\centering
\includegraphics[width=0.85\textwidth]{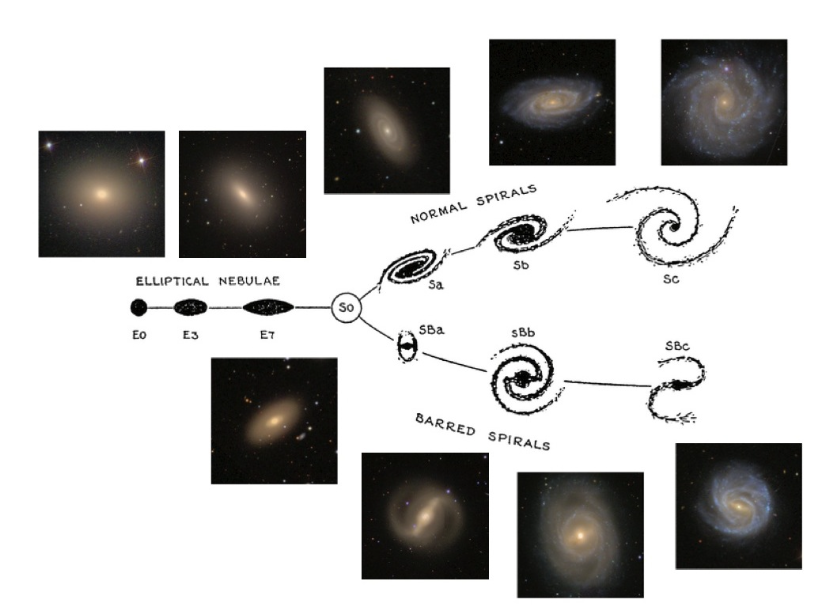}
\caption{An illustration based on a diagram from \citet{Hubble1936} showing the original Tuning Fork classification scheme, with additional example galaxy images from the Sloan Digital Sky Survey (SDSS). Image first published in \citet{Masters2015}.}
\label{fig:tuningfork}
\end{figure}

\subsection{The Development of Classification Schemes}\label{sec:hubble}
 Since classification in science is often the first step to understand a new type of object, it's not a surprise that following (and even before in some cases) the recognition that some ``nebulae" are galaxies external to our own \citep{ShapleyCurtis1921}, many astronomers came up with classification schemes for galaxies.  It is undeniable that the most recognised and used of all galaxy classification schemes is the one based on that first discussed in \citet{Hubble1926}. The basic ``Hubble Tuning Fork"\footnote{A tuning fork is a two pronged fork, which when hit makes a tone useful for tuning musical instruments; in this context it's used to describe the shape of the diagram} structure (see Figure  \ref{fig:tuningfork}) of the Hubble classification scheme with modifications and updates most notable by \citet[][based in part on Hubble's unpublished notes]{Sandage1961}, and notable extensions and refinement in \citet{deVaucouleurs1959} remains ubiquitous in extragalactic astronomy today. 
 %Figure \ref{fig:tuningfork} demonstrates the basic structure. 
 Many authors have written detailed historical accounts of the development and intersections of these various schemes, which I wont repeat here. For further historical information the development of classification schemes the reader may enjoy \citet{Sandage1975,binney1998galactic,Buta2007,Buta2013,Conselice2014,Holwerda2021}. 

As a result of this development, and further refinements, there are many overlapping and related terminologies for galaxy morphology. One of the largest, most comprehensive catalogues of morphologies in common use is the Third Reference Catalogue of Galaxies (RC3). The RC3 was initially published by \citet[][]{RC3}, revised and with errors corrected in \citet{RC3c}. This was an attempt to be a complete catalogue of known galaxies for nearby ($cz<15,000$km~s$^{-1}$) bright and large angular size ($M_B<15.5$ and $D_{25}>1$', where $D_{25}$ is the diameter to the 25th magnitude (in B-band) isophote); it also included galaxies ``of special interest" not meeting those criteria, and/or having been present in the RC2. And importantly it attempted to provide cross comparison of various morphological schemes in use at the time, including ``T-types" a numerical scheme for classifying galaxies which had been developed. 

\subsection{Morphologies in the Era of Big Data Astronomy: Large Surveys and Powerful Computers}
The rise of digital imaging, and projects like the Sloan Digital Sky Survey \citep[SDSS,][]{York2000} and others which initiated ``Large Survey Astronomy" (sometimes ``Big Data Astronomy") caused significant complications for traditional methods of galaxy classification, since by eye inspection of images of all the galaxies in the surveys was a significant undertaking for small numbers of astronomers. When it concluded, the SDSS released imaging for roughly one million galaxies, at least 250,000 of which were bright and large enough in angular scale for detailed morphologies \citep{Willett2013}. While some small teams of professional astronomers still embark on heroic efforts at classifying samples of thousands of these galaxies by eye \citep[e.g.][]{Nair2010catalog,EIFIGBaillard2011,Buta2015}, most moved to alternative, more automated methods. 

\begin{figure}[b]
\centering
\includegraphics[width=0.85\textwidth]{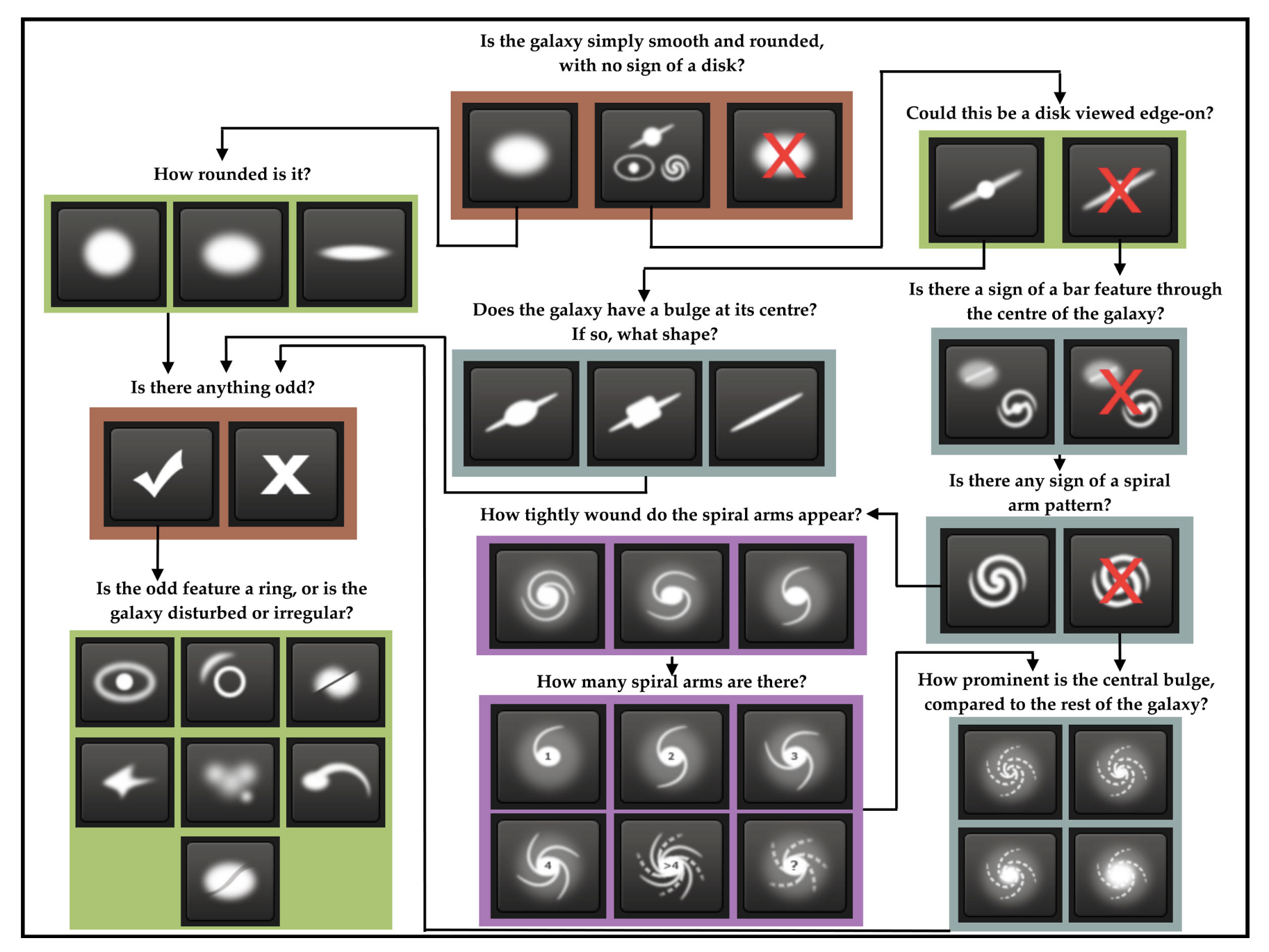}
\caption{A schematic of the classification tree/flowchart used in the second phase of Galaxy Zoo (GZ2), published first in \citet{Willett2013}. This shows the icons used to help describe features to volunteers. Detailed classification trees for each phase of Galaxy Zoo are available at {\tt https://data.galaxyzoo.org/gz\_trees/gz\_trees.html}}
\label{fig:GZtree}
\end{figure}

Morphometric measurements, can be considered excellent proxies for morphology and are well suited to run on large survey data. Computer algorithms can in a semi-automated way extract quantities such as surface brightness profile shapes, total flux in various component, and the ratio between them from images (e.g. the bulge-total flux ratio; one of the largest implementations of which was \citealt{Simard2011}). I call these methods ``semi-automated" as they often require visual inspection for quality control. Non-parameteric measurements on the pixel values in features (e.g. concentration, asymmetry, clumpiness, or CAS \citep{Conselice2003}, or calculating the Gini-coefficient\footnote{The Gini-coefficient is a measure of inequity in a set of data. Readers may be more familiar with its use in economics and inequity in wealth distribution. Applied to images of galaxies, it quantifies how much of the light is in concentrated regions, and how much is spread out. A completely uniformly illuminated image should have a Gini coefficient, $G=0$, while an image with all light in just one pixel, would have $G=1$} of the brightness of pixels) have also been put to use. Another very simple technique, has been to use astronomical colours (or colour index, e.g. $g-r$), which for a long time had been known to correlate well with morphological types. The introductory material in \citet{Smethurst2022} provides a summary of works using galaxy colour as a proxy for morphology (along with advice on when this approximation does/does not work). 

One highly successful method to generate morphological classifications for large samples of galaxies is crowd-sourcing (or ``citizen science"), e.g. as first done by Galaxy Zoo \citep{Lintott2008}. Galaxy Zoo is an internet based project\footnote{www.galaxyzoo.org} which invites members of the public and professional astronomers alike to collaborate in classifying large samples of galaxies. Through a method which combines classifications from large numbers of volunteers highly accurate visual classifications, with estimates of classification certainty are generated. The Galaxy Zoo method uses a ``classification tree" or a series of simple questions was based on identifying a specific feature or features based on a simple description (often "yes/no" for the presence of a feature: see Figure \ref{fig:GZtree} for an example), Galaxy Zoo has been able to collect and publish reliable quantitative classifications for millions of galaxies. The project started with the entire 1 million galaxies in the first SDSS sample \citep{Lintott2011}, moving on in a second phase to provide more detailed classifications for the brightest 250,000 \citep{Willett2013}; the method has also been applied to several other surveys \citep[e.g. see][]{Willett2017,Simmons2017,Walmsley2022,Walmsley2023,Masters2024}. Galaxy Zoo provides morphological classifications as a series of calibrated ``vote fractions" which correlate with how well a given feature is visible in the image of the galaxy. These vote fractions can be used to reconstruct traditional morphological categories if desired (e.g. see the Appendix of \citet{Willett2013} which finds the best T-types for 250,000 galaxies based on Galaxy Zoo classifications), or directly in other ways. 

Another technique which has risen significantly in popularity in recent years, is to apply the methods of machine learning (ML) to classify galaxies. The proliferation of such research makes it impossible to cite all examples; for a recent extensive review on the applications of ML to galaxy surveys (including for morphological classification) see \citet{Huertas-Company2023}. ML is a catch-all term used to describe a suite of methods of computer algorithms which are able to test for patterns in data. The data used for morphological classifications of galaxies could be the images of galaxies directly, or a set of measurements (or morphometric parameters) made on such images. ML algorithms are either supervised or unsupervised. Supervised ML are computer algorithms which seek patterns in data linked to specific previously existing labels - in the case of galaxies, such an algorithm might be taught to recognize images of spirals and elliptical galaxies by being fed a training set of previously classified images \citep[e.g. Zoobot;][]{Walmsley2022}. Unsupervised algorithms will seek patterns in data and cluster items with similar patterns \citep[for a review of such methods applied in astronomy, see][]{Fotopoulou2024}. The pattern clusters generated by an unsupervised ML would then need to be inspected to understand the physical meaning (e.g. all edge-on spirals might be clustered together, so this cluster could later be labelled as edge-on spirals). 

\subsection{A Plethora of Terminologies for Galaxy Types}
 The result of the development of multiple techniques to classify galaxies has been a collection of interrelated, but not always identical in meaning, terminologies in use for galaxy morphology. Table \ref{chap1:tab1}, which was inspired by a similar very useful table published in the RC3 \citep{RC3,RC3c}, provides an updated summary of many of the different terminologies for galaxy classifications in common use today and (roughly) how they correspond to each other. 
 %The table also includes qualitative details of how some morphometric measurements (like the Sersic index, bulge-to-total (B/T) ratio), or photometric or physical properties of galaxies (like galaxy colour, starformation properties or mass) correlate with to these morphologies. 

% \usepackage{multirow}
% \usepackage[normalem]{ulem}
\begin{table}[]
\TBL{\caption{Nomenclature of Galaxy Morphology}\label{chap1:tab1}}
{\begin{tabular*}{\textwidth}{@{\extracolsep{\fill}}@{}lllll@{}}
\toprule
Main Type  & Sub-type   & T-type & RC3 Type & Galaxy Zoo\footnotemark{a} \\
\colrule
\multirow{3}{*}{Elliptical (Early-type, E)} & Compact   & -6    & cE    & 
\multirow{3}{*}{$\left.\begin{array}{l}
                \\
                \\
                \\
                \end{array}\right\rbrace
                \begin{array}{l} 
                \text{Smooth (Elliptical in GZ1)} \\
                \text{Er (round), Ei (inbetween), Ec (cigar-shaped)} \\
                 \end{array}$
                } \\ 
& Elliptical 0-6  & -5     & E0 ... E6 &   \\ 
& cD              & -4     & E+        &  \\
\colrule
\multirow{4}{*}{Lenticular (Early-type, S)}   & -  & 0     & S0   &   
\multirow{4}{*}{$\left.\begin{array}{l}
                \\
                \\
                 \\
                 \\
                \end{array}\right\rbrace
                \begin{array}{l} 
                \text{Smooth (Elliptical in GZ1)} \\
                \text{(unless edge-on,  barred or with a dust-lane,} \\
                \text{then may be featured)} \\
                 \end{array}$
                } \\  
 & Early                      & -3     & S0- &    \\
  & Intermediate               & -2     & S0o  &   \\
    & Late                       & -1     & S0+   \\
\colrule
Spiral (``Late-types"):   &       &        & S                          & Featured, visible spirals                                      \\
   & 0/a                        & 0      & S0/a                       &        \\
          Early-type spirals                                        & a                          & 1      & Sa                         &   Dominant bulge/tight spirals\footnotemark{b}  \\
     & ab                         & 2      & Sab                        &   \\
        Intermediate-type spirals                                          &      & 3      & Sb                         & Obvious bulge/medium spiral\footnotemark{b}   \\
      & bc                         & 4      & Sbc                        &      \\
{Late-type spirals}                & c                          & 5      & Sc                         & Just visible bulge/loose spiral\footnotemark{b}     \\
   & cd                         & 6      & Scd                        &                                                                \\
 & d & 7 & Sd & \multirow{3}{*}{$\left.\begin{array}{l}
                \\
                \\
                \\
                \end{array}\right\rbrace$ No bulge\footnotemark{b}} \\
     & dm                         & 8      & Sdm                        &          \\                                           
Magellanic spiral                                 & m                          & 9      & Sm                         &    \\
\colrule
\multirow{3}{*}{Irregular} & Magellanic                 & 10     & Im        &  \multirow{3}{*}{$\left.\begin{array}{l}
                \\
                \\
                \\
                \end{array}\right\rbrace$ Irregular (under ``odd" features)}  \\
& Compact                    & 11     & cI        &   \\
& Non-Magellanic             & 90     & I0        &   \\ 
\colrule
Peculiar     & (e.g. mergers and more)     & 99     & Pec      & Merger (under ``odd" features)   \\
& & & & Also under ``odd": disturbed, other\\
\colrule
\colrule
\multicolumn{4}{l}{Additional modifiers:}                                                                &                                                                \\
Spirals: & Flocculent & & & Many-arms/can't tell\\
& Multi-arm & & & 4-arms/? \\
& Grand-design & & & 2-arms\\
 \colrule
{Spirals, lenticular or irregular}:  & Un-barred                  &        & A                          & Unbarred                                                       \\
     & Weakly barred (or "mixed") &        & AB                         & Weakly barred\footnotemark{c}  \\
   & Barred                     &        & B                          & Barred                                                         \\
  & Inner ring                 &        & (r)   &    Ring (under ``odd" features)   \\
 & Spiral (or S-) shaped      &        & (s)                        &  Visible spiral \\
 & Ring+spiral (Mixed)        &        & (rs)             & Ring (under ``odd" features) \\
& Nuclear ring               &        & (nr)                       & --  \\
\colrule
All types:   & Dwarf   &        & d   &  --  \\
           & Uncertain                  &        & :                          &                                                       --         \\
               & Doubtful                   &        & ?                          &                      --        \\
         & Spindle (edge-on)          &               & sp                         &      Edge-on/cigar shaped smooth  (Se or Ec)                                                      \\
     & Outer ring                 &        & (R)   &   \multirow{2}{*}{Ring (under ``odd" features)} \\
       & Pseudo-outer ring          &        & (R')    &   \\  
\colrule
\end{tabular*}}{%
\begin{tablenotes}
\footnotetext[a]{As best matched}
\footnotetext[b]{This combination is the traditional Hubbles spiral-type, but note \citet{Masters2019} could find no strong correlation between arm winding and bulge size in GZ2 classifications.}
\footnotetext[c] Label present in some versions of Galaxy Zoo, but not all. 
\footnotetext{\source{Table based in part on Table 2, Section 3.3.a of the RC3 \citep{RC3,RC3c}, see {\tt https://heasarc.gsfc.nasa.gov/w3browse/all/rc3.html}.}}
\end{tablenotes}
}
\end{table}

\section{Types of Galaxies}\label{sec:traditional}
What follows is a summary of the morphological, and physical features found in many types of galaxies in the Universe, along with comments on the intersections between different terminologies.

 \subsection{Spiral and/or Disc Galaxies}

\begin{figure}[b]
\centering
\includegraphics[width=0.6\textwidth]{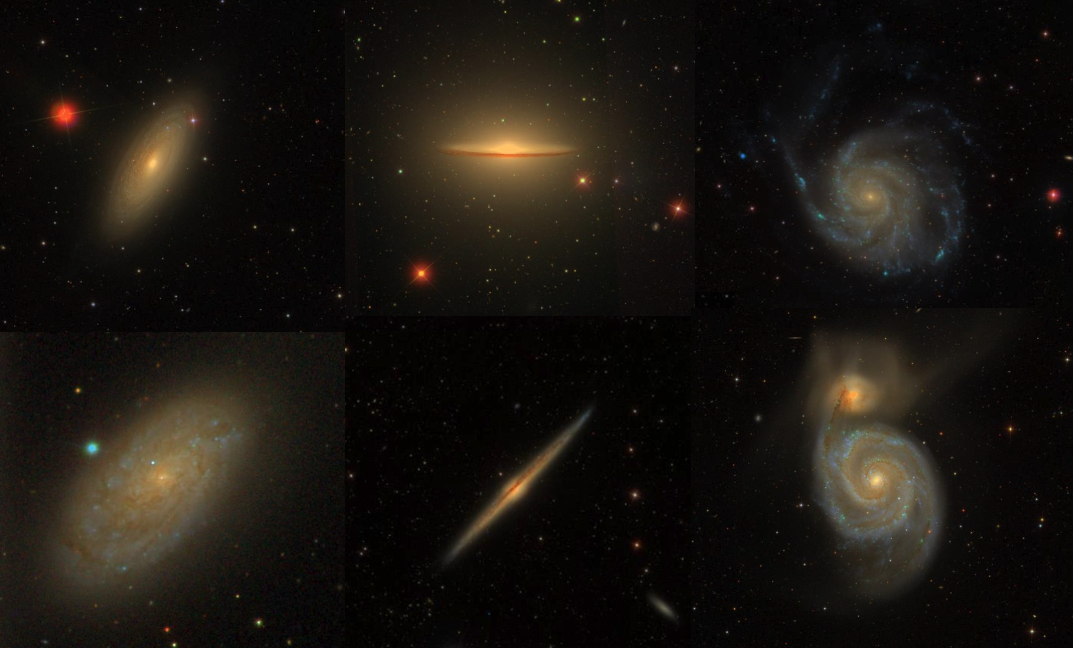}
\caption{Six very nearby galaxy examples illustrating some of the variety of spiral/discs seen even in unbarred spirals. All images are SDSS $gri$ composites. Shown are: {\bf Left column:} two flocculent spirals: NGC 2841 (upper)  which is usual called an Sa today, but was noted as an example of Sb by \citep{Hubble1926} and NGC 4298 (lower) an SA(rs)c which was classifed in GZ2 with vote fractions $p_{\rm features} = 0.978$, $p_{\rm no~bar} = 0.837$, $p_{\rm spiral} = 0.884$, $p_{\rm tight~spiral} = 0.895$, $p_{\rm can’t~ tell~N-arms} = 0.711$. {\bf Middle column}: two edge on discs (NGC 4594 (The Sombrero) proto-typical SA(s)a (example from Hubble’s 1926 paper) and NGC 4565 (The Needle Galaxy) SA(s)b. {\bf Right column}: Two face spirals, upper: Multi-arm spiral NGC 5457 (M101 or The Pinwheel) an SAB(rs)cd and lower: grand design spiral spiral NGC5195 M51 (Whirlpool) SA(s)bc pec. }
\label{fig:spirals}
\end{figure} 

  The most common type of massive galaxy in our Universe - spiral galaxies are easily identified by their spiral arms, features which extend from the central regions of the galaxy in a number of spiral shapes (most often two, but sometimes more or less). Indeed spiral arms in galaxies were noted well before we understood what galaxies are \citep{Rosse1850}. Within the broad class of spirals there is great variety, as illustrated in Figures \ref{fig:spirals},\ref{fig:bar},\ref{fig:rings}. The three-dimensional structure of a spiral galaxy is that of a flattened disc, with the motion of stars dominated by ordered rotation.  As a result spiral galaxies are sometimes lumped into a broader ``disc galaxy" category (which would also include lenticular, or S0) galaxies; see Section \ref{sec:S0} below). 

The surface brightness (or light) profile of a galaxy describes how the aximuthally averaged surface brightness changes with the semi-major axis, or elliptical radius of the ellipse it is calculated in. A rotating disc of stars under gravity naturally generates an exponential disc, or a surface brightness profile of $I(r) \propto \exp{(-r/r_s)}$, where $r_s$ is the scale length; for surface brightnesses measured in magnitudes this is a linear decline with radius. Another common functional form used for light profiles is the Sersic profile \citep[$I(r) \propto \exp{(-k r^{1/n_s}})$, ][]{Sersic1963}. The discs of spirals have $n_s=1$ Sersic profiles. The Sersic profile is commonly expressed in terms of $r_e$, or the effective radius (radius containing half the light) and $I_e$ (the surface brightness at the effective radius) as
\begin{equation}
    I(r)  = I_e \exp \left\{ -b_n \left[ \left(\frac{r}{r_e}\right)^{1/n_s} -1  \right] \right\}, \label{eqn:sersic}
\end{equation}
where $b_n(n_s)$ is a fit parameter found so that $r_e$ is the radius in which half the total light is contained. 

Many spiral galaxies, including our own Milky Way, contain a central ``bulge" of more concentrated light, which typically also has a spheroidal shape, extending above and below the disc, and in some cases is better fit by an $n_s=4$ Sersic profile (also known as a de Vaucouleurs profile; see Section \ref{sec:ellipticals}), although ``disky" bulges (or $n_s=1$ ``pseudo-bulges") have also been found \citep{Kormendy2004}. 

\subsubsection{Types of Spirals}
The Hubble spiral sequence \citep[][see Figure \ref{fig:tuningfork}]{Hubble1926} placed spirals in a three category sequence from {\bf Sa} (closest in form to the elliptical galaxies), {\bf Sb} to {\bf Sc}. \citet{deVaucouleurs1959} added intermediate types (like {\bf Sab} etc) and extended the sequence to include types {\bf Sd} through to {\bf Sm} (for LMC like spirals), which Hubble had called ``Irregular". The Hubble spiral sequence was initially set up using a combination of bulge size and spiral arm morphology which \citet{Hubble1926} described as being based on: 
\begin{enumerate}
    \item ``relative size of the unresolved nuclear region" (aka bulge size)
    \item ``extend to which the arms are unwound" (aka pitch angle of the spiral features, as described below)
    \item ``degree of resolution in the arms" - which might be thought of as how smooth the arms are (presumably highly related to the classification of ``flocculent" or ``grand-design", see Section \ref{sec:flocc}). 
\end{enumerate}
\citet{Hubble1936}, reflecting on his classification scheme (and defending it to criticism by \citealt{Reynolds1927}) reflected that "among upward of a thousand spirals which I have examined, not more than a dozen have refused to fit in the sequence", and it is perhaps its simplicity (and extensions) that have led to it continued use to today. 

Recent analysis suggests that in modern use, the placing of spirals along the Sa--Sc sequence is primarily driven by bulge-disk size or luminosity ratio \citep[$B/T$; e.g.][]{Willett2013,Masters2019}. It is also interesting to note that while there are correlations between bulge size and pitch angle in some samples of galaxies \citep[][particularly the bluer, lower mass spirals that would have been characteristic of Hubble's sample on photographic plates]{kennicutt1981,Mengistu2023} in other samples including more massive spirals, no correlation can be found \citep{Masters2019,Mengistu2023}. 

Spiral galaxies are also often referred to as ``late-type" galaxies (or LTGs) and spirals with smaller bulges as more ``late-type" spirals (to contrast with the ``early-type" galaxies, or ETGs being ellipticals and lenticulars; see Section \ref{sec:ellipticals}). The author has also noted this ``LTG/ETG" distinction being used to classify galaxies on the basis of their star formation properties rather than morphology, with ``late-type" used to mean galaxies actively forming stars. Of course, it is true that most spiral galaxies are actively forming stars, and as such appear optically blue \citep{Strateva2001}, but there exist a small, but non-negligible fraction of spirals which are relatively quiescent in their star formation properties \citep[e.g.][]{Masters2010red} appearing optically red, and in addition dusty star-forming spirals, particularly viewed edge-on, will also appear optically red. Selection of spirals by colour (or SF properties) will result in a sample which both misses these red objects and contains a small number of star-forming, or blue elliptical galaxies \citep{Smethurst2022}.

There is a persistent astronomy urban legend that says Hubble invented the early-type/late-type terminology for galaxies to imply an evolutionary timeline for galaxies from ellipticals to spirals, however if you return to the original use in \citet{Hubble1926} you will find the following statement: ``temporal connotations are made at one’s peril”, and further text which explains that the terminology was based on analogy with classification nomenclature of stars in use at the time. \footnote{``Early-type" stars was a term historically used by astronomers to mean the hotter, OB and maybe A-type stars, while "late-type" stars refered to cooler stars. This terminology was a hold-over from early models of stars being powered by gravitational contractions; which we now know not to be true.}

\subsubsection{Spiral Arm Morphology}\label{sec:flocc}
 Spiral galaxies can have different numbers of spiral arms, and also the shape of those spirals can vary from almost circular features, hard to distinguish from rings (see Section \ref{sec:rings}), to very open linear features, hard to distinguish from tidal tails. A spirals shape, is usually characterized by a pitch angle, defined as the angle of the arm to the tangent of a circle at the same radius. So a circle has a pitch angle of zero; radial spokes have a pitch angle of 90$^\circ$. Since the time of \citet{Hubble1926} astronomers have also considered spiral ``winding", as a more qualitative description of how tightly-wound spiral features are (from tightly wound, aka small pitch angle, to loosely wound, aka large pitch angle); Galaxy Zoo asks questions about spiral galaxy features using this terminology. 
 
 There's no clear reason why spiral arm pitch angles should be constant with radius, but it is generally observed that spiral arms in galaxies have shapes surprisingly close to log-spirals; a mathematical description of a type of spiral, common in nature, in which the pitch angle does not vary with radius. In polar co-ordinates log-spirals are described as $r = a\exp^{k\theta}$, where the pitch-angle, $\phi$ is found from $\tan \phi = k$. It is observed that there can be some variation in pitch angle between different arms in a single galaxy (e.g. $\Delta \phi = 11\pm1^\circ$ was found by \citealt{Lingard2021}) and pitch angles are notoriously tricky to measure, due to the patchy, and sometimes varying widths of spiral features in real galaxies, meaning different methods also sometimes result in wildly differing pitch angles for the same galaxy. Never-the-less pitch angles of spirals should provide fascinating constraint on spiral formation mechanisms \citep{Sellwood2022}, and dynamics of the host galaxy, so hopefully continued effort will reveal more robust methods to determine pitch angles. 

 Another property of spiral arms, first described by \citet{Elmegreen1982}, and not always captured clearly in other classification schemes is referred to as the ``arm class". Initially set out as an 12 category classification scheme correlating with how ``orderly and symmetric" the spiral arms are, now the most used version separated spiral arm morphologies into the following three categories: 
 \begin{itemize}
     \item {\bf ``flocculent"} - lacking bimodal symmetry and composed only of small pieces (``flocculent" literally meaning ``fleece like" or fluffy)
     \item {\bf ``grand design"} - two symmetric arms (at least over most of the galaxy), which are continuous across a large range of radii
      \item {\bf ``multi-arm" or ``intermediate"} - show some features of both flocculent and grand design, typically having more than two arms, which may be continuous over some or most of the galaxy 
 \end{itemize} 
 Flocculent spirals tend to be lower mass than grand design, and curiously many studies include only these two types, considering ``multi-arm" spirals as just intermediate. However the physical properties of ``multi-arm" may suggest other wise \citep[e.g.][who show they tend to be even more massive than grand design spirals]{Smith2024}.

 \subsubsection{Galactic Bars}
\begin{figure}[b]
\centering
\includegraphics[width=0.4\textwidth]{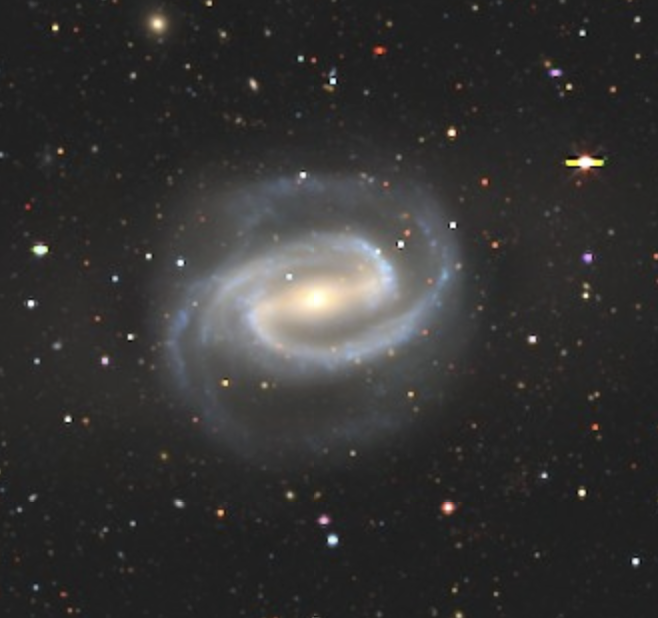}
\caption{An image of the strongly barred galaxy, NGC 1300 taken as part of the DESI Legancy Imaging Survey.}
\label{fig:bar}
\end{figure} 

Galactic bars (presumably named by their resemblance to iron bars) are linear features stretching across the center of many spiral (or disc) galaxies. Similar in some ways to the spiral arms, a bar is a dynamical feature which naturally develops in a disc of orbiting stars under the right combination of rotational and vertical motions and having to do with the general stability of the disc. Bars themselves have a variety of morphologies from ``weak" bars (``SA" in the Hubble terminology) which at times are only mild oval distortions and can be hard to distinguish from bulges, to ``strong" bars (``SB" types) which may appears to dominate the disc (e.g. the prototype being NGC 1300, see Figure \ref{fig:bar}). In samples of galaxies with access to dynamical information, bars may also be characterized by their ``speed" with ``fast" bars being those which have angular pattern speeds, $\Omega_b$, comparable to the angular orbital speed of stars at their ends (because they extend in length to at least 70\% of the ``co-rotation" radius, or have $\mathscr{R} \equiv R_{\rm CR}/R_{\rm bar}$ in the range 1--1.4) and ``slow" bars, which have pattern speeds that mean they are moving much more slowly than the stars at their end (i.e. they are much smaller than the co-rotation radius, or $\mathscr{R}>1.4$). ``Ultra-fast" bars (those which extend beyond co-rotation, or $\mathscr{R}<1$) are not supposed to be dynamically possible, but have been observed \citep[e.g.][]{Garma-Oehmichen2022,Geron2023}, but this may be as a result of the challenges of measuring pattern speeds.

In the RC3 \citep{RC3} the optical bar fraction (fraction of spirals having a bar) is around $f_{\rm bar} \sim 0.25–0.3$, a number comparable to more recent analysis using optical images \citep[e.g.][]{Masters2011,Nair2010}. However the measured bar fraction can rise to 60--75\% if weaker bars or oval distortions are included, and when looking at NIR imaging of galaxies \citep[e.g.][]{Menendez-Delmestre2007,Sheth2008}. It's dynamically curious that any disc galaxy would exist without a bar; observationally this is found to be most common in the lowest mass and most gas rich galaxies, with redder more massive discs (including some non-spiral discs; or lenticulars) having much higher bar fractions, particularly for strong bars \citep[e.g.][]{Masters2011,Geron2021}. Bars have been linked with a whole host of physical processes important for galaxy evolution, including their potential role in feeding gas to active galactic nuclear, and the quenching star formation. They can be found in all types of spirals, including Sm or LMC like, including the LMC itself, which has a fascinating offset bar \citep[e.g.][]{Kruk2017}, as well as lenticular galaxies.

\subsubsection{Nuclear, Inner and Outer Rings}\label{sec:rings}
Rings, or circular closed feaures are found in many disc galaxies, They are mostly interpreted as the locations of various dynamical resonances in galaxies where gas (and as a result starformation and new bright stars) have collected. Rings are particularly notable in galaxies with strong bars, but can occur in other galaxies. The bar (or any large scale pattern in a galaxy), which rotates at a pattern speed, $\Omega_b$, generates various resonances with the orbits of stars, for example co-rotation (the radius at which stars orbit at the same angular speed as the (bar) pattern, or $\Omega_\star 
= \Omega_b$), and Lindblad resonances, where the period of radial oscillations of the stars is some integer fraction of the difference between the angular speed and the pattern speed ($m(\Omega_\star - \Omega_b) = \pm\kappa_\star$, where $\kappa_\star$ describes radial angular oscillation of stars). 

 Observationally, rings come in several main types. {\bf Nuclear} rings (nr) are found only in the very centers of galaxies, {\bf Inner rings}, denoted (r) are typically found at or close to the end of a bar (if present) and {\bf outer rings} (R) are found much further out, often with a distinct gap between them an the outer parts of the main galaxy. Another type of ring is a {\bf collisional ring} (the Cartwheel galaxy being the typical proto-type). Some galaxies, like Hoag's object, appear to be just a ring of starformation in an otherwise low density disc. 
 
\begin{figure}[b]
\centering
\includegraphics[width=0.4\textwidth]{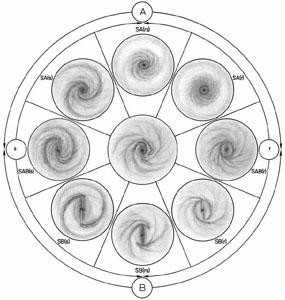}
\caption{An illustration of the different types of Sb galaxies identified by \citet{deVaucouleurs1959} showing barred (B) and unbarred (A), inner ring (r) and spiral (s) and all intermediate types. This image from \citet{Buta2007}. }
\label{fig:rings}
\end{figure}

 For a lot more more details on rings in galaxies, see \citet{Buta1999}; for a large recent ring galaxy catalogue see \citet{Buta2017}. The \citet{deVaucouleurs1959} classifications scheme includes rings as a second dimension in the direction perpendicular to the spiral types Sa-Sc sequence (bar strength being the first),  as illustrated in Figure \ref{fig:rings}.

\subsubsection{Edge-on Discs}
As three dimensional flattened structures, the visual morphology of spirals (or any disk) galaxy depends on viewing angle. Astronomers describe this angle using an inclination, $i$, which is the angle the normal to the disc makes with the line of sight ($i=0^\circ$ being completely face-on, $i=90^\circ$ completely edge-on). 

Galactic discs are very thin relative to their radial extend, but not infinitely thin, so the relationship between the observe axial ratio (ratio between the major ($a$) and minor ($b$) axis) and inclination is given by
\begin{equation}
    \cos^2 i = \frac{ (b/a)^2 - 1 }{1-q^2},\label{eqn:inc}
\end{equation}
where $q$ is the intrinsic axial ratio, or the axial ratio of the galaxy if it were viewed completely edge-one. Obviously for a given galaxy (unless viewed completely edge-on) this quantify cannot be known, but population statistics can be used to estimate it for types of spirals finding that earlier type spirals (Sa) and lenticulars are typically thicker than later-type spirals; values of $q=0.14-0.2$ are commonly used. Equation \ref{eqn:inc} also assumes that seen face-on spiral discs are completely round. This can also be tested either in completely face-on samples, or using statistics of population shapes, and is found to be close to true \citep[e.g.][]{Ryden2004}.

Viewed completely edge-on, disc galaxies sometimes called ``spindle" galaxies can show interesting morphological features, such as dust-lanes, and can also be interpreted as a mixture of thinner and thicker discs (the thicker disc typically being redder as it's made of older stars). The shape of the central bulge of an edge-on galaxy can also vary from rounded, to boxy or ``peanut" shaped (sometimes boxy/peanut, or b/p), and even ``X-shaped". These latter shapes are interpreted as revealing a bar viewed end-on.   

 \subsection{Early-types: Elliptical and Lenticular Galaxies} \label{sec:ellipticals}
It's in somewhat common use to say ``early-type" to mean any galaxy which isn't a spiral galaxy. Some works use this ETG almost interchangeably with ``elliptical" galaxy, however strictly, ``early-type" also includes lenticular (or S0) galaxies, and may even include some of the ``earlier" type spiral galaxies (Sas). In \citet{Strateva2001}, the paper most often cited as showing that spirals are blue and ellipticals are red, they actually compared the colours of ETGs (defined in the paper as E+S0+Sa) compared to LTGS (defined as Sb and later). Other astronomers define ETG directly using the star formation properties (i.e. to mean galaxies with star formation rates below that of the typical star forming sequence). As previously noted, while colour/star formation rates and morphology correlate well broadly, the use of colour to select for morphology generates rather mixed samples, particularly when trying to isolate elliptical or lenticular galaxies, since any sample of red galaxies will contain a non-negligible fraction of red spirals \citep{Smethurst2022}. Red spirals are mostly Sas, but even some Scs are quite red/low SFR \citep{Masters2010red}, and many of the reddest galaxies locally are edge-on discs \citep{Sodre2013}.

 \subsubsection{Elliptical Galaxies}
 Elliptical galaxies appear completely smooth and featureless. They are typically denoted by the letter ``E" followed by a number, $n$ which describes their ellipticity (or ratio of minor axis, $b$ to major axis, $a$) via 
 \begin{equation}
 n = 10\left[ 1 - (b/a) \right]
 \end{equation}
 Observationally elliptical galaxies vary from E0 (completely round) to E6, which are highly elongated or ``cigar" like. The three dimensional structure of elliptical galaxies is understood be that of a triaxial spheroid (where the three axes, $a$, $b$, and $c$ all have different values - like a kiwi fruit, which when cut through has an elliptical cross section). They might vary from very close to prolate (sausage shaped) to oblate (pancake shaped). Viewed in projection on the sky, the radially averaged surface brightness profile best fits a de Vaucouleurs profile \citep[$\log I(r) \propto r^{1/4}$; ][]{deVaucouleurs1948}; in the SDSS photometric pipeline these are noted as {\tt frac\_deV} $\sim1$  objects - i.e. in which the de Vaucouleurs profile alone can best fit the profile. Another way to say this, is that they have a Sersic profile with $n_s=4$ (see Equation \ref{eqn:sersic}). This light profile is more highly concentrated that the exponential profile of disc galaxies, hence elliptical galaxies are also more concentrated (the ``C" in ``CAS", or parameterized in other ways) than discs. 

 The textbook interpretation of ellipticals is that they have stellar orbits dominated by random motions, however large surveys of the dynamics of elliptical galaxies reveal that many contain a substantial component of stars rotating in a disc \citep{Emsellem2011}.
 
 Less common, sub-types of ellipticals might be described as cE ({\bf compact ellipticals"}) and cD ({\bf ``central diffuse"}). The term dwarf elliptical (dE) is also used, but confusingly this may actually be a low mass smooth disc galaxy (see Section \ref{sec:dwarfs} for more on this). 
 
\begin{figure}[b]
\centering
\includegraphics[width=0.4\textwidth]{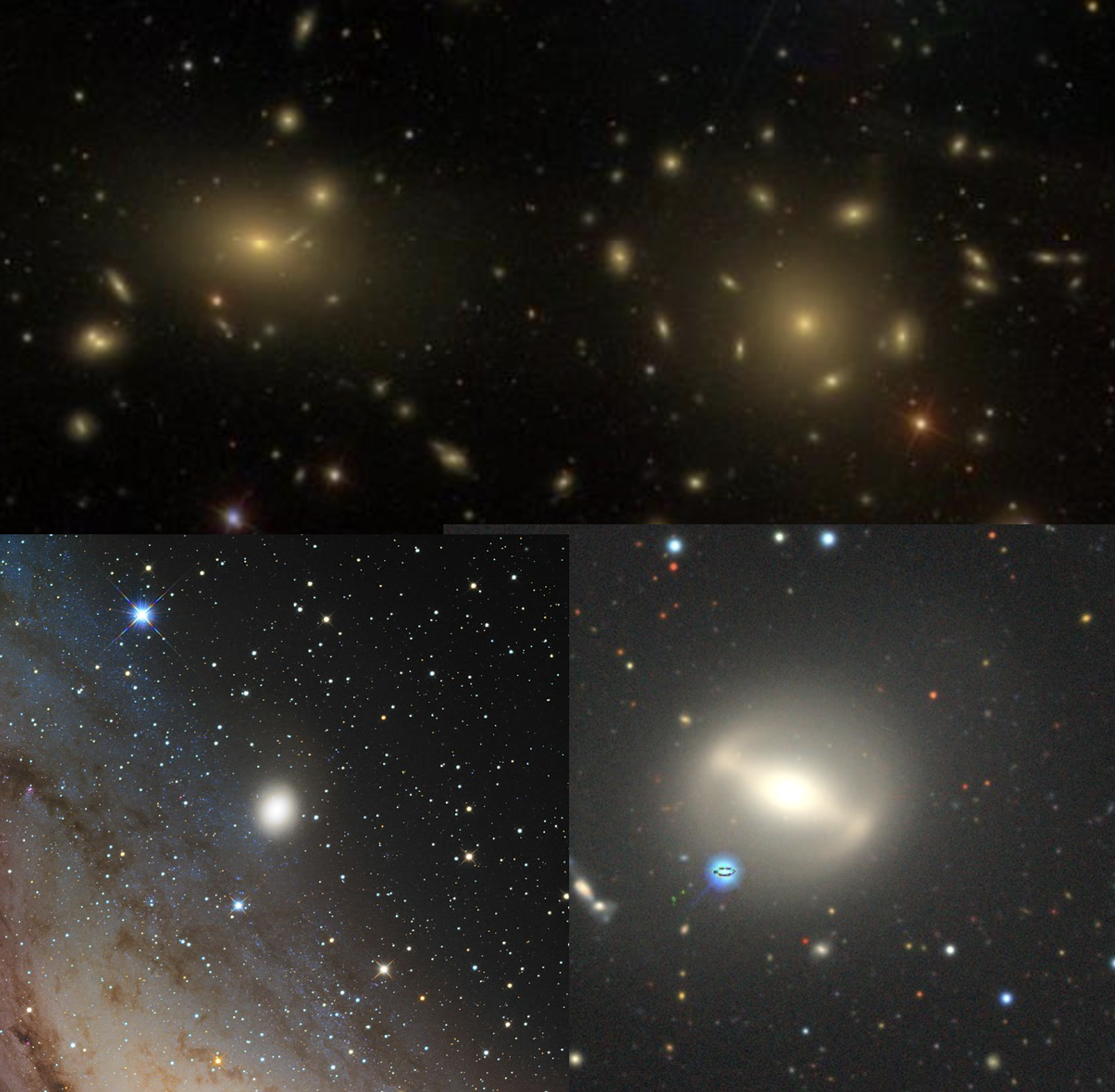}
\caption{An illustration of some of the different kinds of early-type galaxies. Upper: NGC 4874 and NGC 4889, the joint BCGs, or types of cDs at the center of the Coma cluster. Lower left: M32, a cE galaxy in an image from a small telescope taken by Fabrizio Francione. Lower right: the barred S0 galaxy NGC 1460 in DESI imaging. }
\label{fig:Es}
\end{figure}

  The notation cD was originally used to mean a giant (c), diffuse (D) galaxy (part of a now little used classification scheme of \citealt{Morgan1958}), but may be easier remembered as meaning "central dominant" galaxies since cD ellipticals (sometimes called E+ galaxies) are the giant ellipticals at the center of many galaxy clusters (a significant fraction of brightest cluster galaxies, or BCGs are cDs). cDs are extremely massive, show a compact core and a diffuse lower surface brightness halo taken to reveal material stripped off other galaxies in the cluster. Examples include NGC4874, and NGC4889 the joint BCGs in the Coma cluster (see Figure \ref{fig:Es}).

  An even rarer type of elliptical is a {\bf compact elliptical} (cE). These are much lower mass (M32; a satellite of the Andromeda galaxy is a proto-type, see Figure \ref{fig:Es}), although too massive to be considered dwarf galaxies, and are distinguished by their much more compact core than a typical elliptical. There are some suggestions that this type reveals the stripped bulge of what was previously a spiral or disc galaxy, although a significant fraction are found in isolated environments \citep{Deeley2023}. 

\subsubsection{Lenticular, S0 Galaxies}\label{sec:S0}
 Lenticular galaxies (denoted S0) may also appear smooth in images, and their central regions in particular may strongly resemble an elliptical galaxy.  However the outer regions and surface brightness profile will the presence of a (three dimensionally) flat disc (e.g. by being best fit by a Sersic with $n_s<4$) and edge-on lenticular galaxies can be almost indistinguishable from edge-on spiral galaxies. Lenticular galaxies may also contain a (smooth) bar (see Figure \ref{fig:Es} for an example of a barred lenticular). The S0 classification was initially added as an intermediate between ellipticals and spirals, and notations  S0-, S0$^\circ$ and S0+ may be used to indicate a sequence from most elliptical like, to most disc like.  

 Most lenticular galaxies show evidence for a substantial rotating disc. Many papers have been written about their formation; faded spirals, or spirals stripped (of gas) or experiencing a merger which removed (or uses up) the gas while leaving behind a rotating disc being common suggestions \citep[e.g. the introduction of][provides a nice review]{Deeley2021}. 

\subsection{Peculiar or Irregular Galaxies}
Galaxies can look unusual (or peculiar, or irregular) in a variety of different ways. Mostly we mean they are not symmetric - neither symmetric spirals, nor symmetric and smooth, although to some extent, most galaxies are slightly asymmetric; peculiar or irregular galaxies are usually significantly asymmetric. Since the definition of such objects is that they have no clear regular defining structures, it's a bit hard to describe what they look like generally. However they are often low mass, highly star forming and blue/patchy in structure.

Hubble called any galaxy which didn't fit his classification scheme ```Irregular", separating them into ``Type I" and ``Type II" irregulars. Hubble's Type I irregulars would later  be reclassifed as types of low mass, LMC, or dwarf-like spirals, or Sd, and Sm \citep[starting with ][]{deVaucouleurs1959}. In some cases faint spiral structure is visible in these, but it is much more irregular in the more massive spirals. 

The Type II Irregulars (Irr II) are more irregular still, but are notable for having no obvious tidal features (so not fitting into any merger sequence as described below). In any classification scheme of galaxies there will be some outliers, and this is the classification for those. In modern surveys including millions of galaxies there are in fact many thousands of such objects, and they can provide fascinating insights into galaxy evolution.

\subsubsection{Merging Galaxies}
\begin{figure}[b]
\centering
\includegraphics[width=0.7\textwidth]{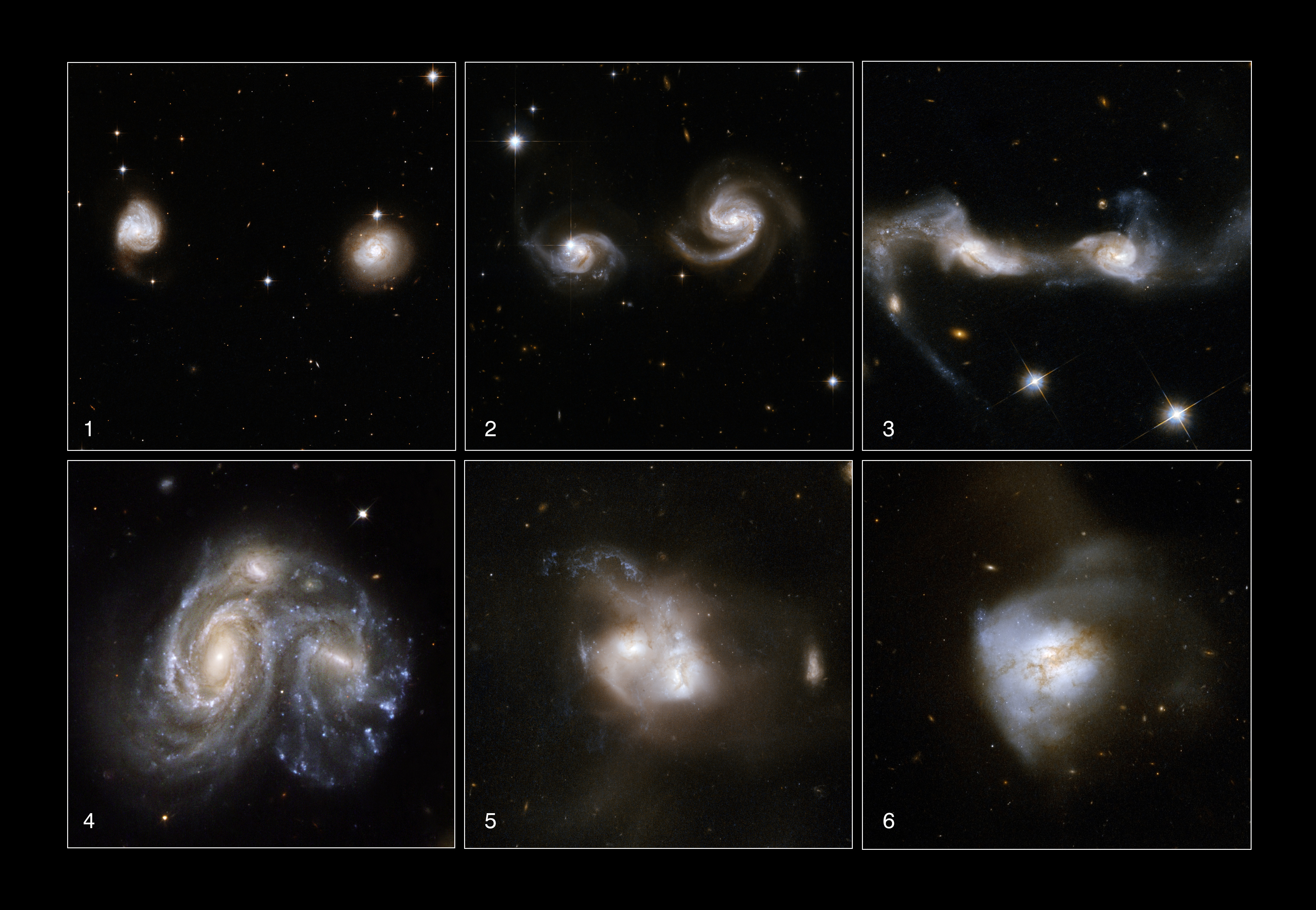}
\caption{A series of six HST images of galaxies ordered in an attempt to make a merger sequence from the first hints of an interaction (upper left), to a post-merger irregular object (lower right). Credit: NASA, ESA, the Hubble Heritage Team (STScI/AURA)-ESA/Hubble Collaboration and A. Evans (University of Virginia, Charlottesville/NRAO/Stony Brook University), K. Noll (STScI), and J. Westphal (Caltech) }
\label{fig:mergers}
\end{figure}

A special case of peculiar or irregular galaxies are those which can be identified as being due to mergers or tidal interactions. When galaxies gravitationally interact, they can be distorted into a wide variety of unusual morphologies including tidal bridges and streams and a variety of other structures. One of the first catalogues of such objects was the ``Atlas of Peculiar Galaxies" by \citet{Arp1966}. In many cases it's possible to make out the original morphological type of the two (or more) galaxies in the system and so they may be classified as such. Attempts are also made to place such objects into merger sequences (e.g. Figure \ref{fig:mergers}). 
Many different morphometric parameters have also been developed to identify merging systems, such as the asymmetry criteria of the CAS \citep[Concentration-Asymmetry-clumpinesS][]{Conselice2003} system, or RMS (root-mean-square) Asymmetry \citep{Sazonova2024} which attempts to also quantify the impacts of imaging depth on the visibility of asymmetry (since deeper imaging almost always reveals more asymmetric features in galaxies).

\subsubsection{Active Galactic Nuclei}
We don't often consider morphology as a primary characteristic of galaxies with active galactic nuclei (or AGN), but of course some fraction of such objects show a bright point source not only in X-ray or radio, but also in the optical. The most extreme example of this are the ``quasars" or ``quasi-stellar objects" (QSOs) so named because their visual appearance mimics that of a star - they show no extended structure (at least in ground based imaging). Even let bright AGN can generate a point source, and one of the delightful surprises in recent JWST imaging of distant galaxies has been just how many showed the six spiked diffraction pattern revealing an unresolved point source at their nucleus.  

\subsubsection{Low Surface Brightness Galaxies and/or Ultra Diffuse Galaxies}
Most of what has been written about galaxy morphology is assuming fairly standard depth imaging. Much deeper imaging reveals around almost any galaxy a wide variety of low surface brightness (LSB) features, and suggesting that as imaging surveys improve both in resolution and depth, more and exciting morphologies of galaxies will be reveals. Some galaxies in their entirity are very low surface brightness. The discovery of Malin 1 \citep[][]{Bothun1987} started to reveal this LSB universe. Many LSB galaxies are dwarf galaxies (and many dwarf galaxies are LSB), but {\bf Giant Low Surface Brightness Galaxies} (GLSBG), like Malin 1 are some of the most massive spirals we know of. Sometimes they were misclassified as ellipticals in less sensitive imaging, but deeper imaging revealed faint spiral structure.  Recently, even more diffuse galaxies, dubbed {\bf Ultra Diffuse Galaxies} (UDG) have been noted in very deep wide area imaging of the Coma cluster \citep{vanDokkum2015}. Most are noted as extremely red and round, and with very high dark matter fractions, although the UDG, NGC1052-DF2 has been reported to completely lack dark matter \citep{vanDokkum2018}.

 \subsection{Dwarf Galaxies} \label{sec:dwarfs}
By sheer number, most of the galaxies in the Universe, are dwarf galaxies, defined as any galaxy smaller than a certain brightness or stellar mass limit (the exact definition may vary, something like $M_\star<10^8 M_\odot$, or $M_B > -18$ is common). The Milky Way is surrounded by a substantial number of dwarf galaxies. And while we assume this property is common to most massive galaxies, due to their small size, most observations of dwarf galaxies are from just the very local Universe. 

There are many different types of dwarf galaxy morphology you may find in use in the astronomical literature. Usually they are named as dwarf versions of common morphological types. What follows is a short summary of the defining characteristics of some of the more common terminologies: 

\begin{itemize}
\item{\bf Dwarf Elliptical, dE}: a galaxy which is elliptical in features, but much smaller than either normal ellipticals (E) or compact ellitpicals (cE). Such galaxies are also typically much bluer than more massive ellipticals, and have a more exponential surface brightness profile. These objects may be example of primordial galaxies, or stripped spirals. Example: NGC 205, NGC 147, NGC 185.
\item{\bf Dwarf Spheroidal, dSph}: a galaxy typically even smaller than a dwarf elliptical, low in luminosity and surface brightness, and typically only hosting older stars. Examples: NGC 147, NGC 185 (but these are sometimes both classified as dE), and many of the dwarf galaxies in the local group. 
\item{\bf Dwarf Irregular, dI}: a dwarf galaxy with an irregular structure. Example: UGC 4459. 
\item{\bf Dwarf Spiral, dS}: a dwarf galaxy with a spiral structure. Such objects are very rare. Example: NGC 3928.
\item{\bf LMC Type Dwarfs, dm}: named for their proto-type the LMC, these are dwarf versions of LMC type spiral galaxies (Sm) a type of irregular galaxy which shows some spiral structure. Example: NGC 5474. 
\item{\bf Blue Compact Dwarfs}: A type of compact dwarf galaxy which is very blue in colour. They might also be categorized as one of the other dwarf types (e.g. dIrr). Example: NGC 1705.
\item{\bf Ultra-Compact Dwarfs, UCD and Green Peas}: names for very compact dwarf galaxies. UCDs are particularly high density compact dwarfs. Green Peas were named by Galaxy Zoo volunteers \citep{Cardamone2009} for their compact round green appearance in SDSS imaging. 
\end{itemize}

%\begin{align}\label{chap1:eq1}
%p[m_1]+\cdots+p[m_2]=p[n_1]+\cdots+p[n_2]
%\end{align}

%\begin{align}\label{chap1:eq2}
%\mathrm{{H_{2}}^{+}} + \mathrm{e}^{-} & \rightarrow  %\mathrm{H} + \mathrm{H}, \\
%\mathrm{HeH^{+}} + \mathrm{e}^{-} & \rightarrow  %\mathrm{He} + \mathrm{H}.\label{chap1:eq3}
%\end{align}

%\include{table_example}

\subsection{Morphologies of Galaxies in the Early Universe} 
In the era of large space telescopes (first HST, launched in 1990 and then JWST in 2021) bringing sub-arcsecond resolution to optical imaging it has been possible to consider the morphology of galaxies at very high redshifts, thus much earlier in the history of galaxy evolution. This work has revealed a universe of galaxies with some similarities, but many differences to the more familiar local Universe \citep[e.g.][]{Elmegreen2004}. Merging and interaction were clearly more common in the early lives of galaxies, and many high redshift galaxies show much more peculiar, clumpy structures, likely revealing the sites of active star formation. One challenge has come from the redshifting of light - high redshift galaxies viewed in the optical are seen in rest-frame UV, where even local galaxies can look much clumpier. The launch of JWST and it's longer wavelength imaging is in the process of mitigating this, and much new fascinating galaxy dynamics and evolution is sure to be revealed from such work.

%\section{Why Galaxy Morphology is Useful} 

\label{sec:highz}

%\begin{itemize}%
%\item The atmosphere is dry with no phase changes of water occurring.
%\item Comparatively short time periods are involved so that radiational
%\end{itemize}

%\begin{enumerate}%
%\item The atmosphere is dry with no phase changes of water occurring.
%\item Comparatively short time periods are involved so that radiational heating or cooling of the air is relatively small.
%\begin{enumerate}[a.]%
%\item conservation of motion,
%\item conservation of water, and
%\end{enumerate}%%
%\end{enumerate}%%

%\begin{description}
%\item[A setup-time $T_c$ per pair of nuclear centers.] For interelectron repulsion
%\end{description}

%\begin{quote}
%\quotehead{Quotehead}
%The foundation for any model is a set of conservation principles. For mesoscale atmospheric models, these principles are conservation of mass, conservation of heat, conservation of motion, conservation of water, the conservation of other gaseous and aerosol materials, and an equation of state.
%\source{--source}
%\end{quote}

\section{Conclusions}\label{sec:ML}%
The morphological classification of galaxies is a science with a long history and a bright future. The physical information which can be found by looking at images of galaxies, at an ever increasing variety of resolutions, depths and wavelengths can provide a wealth of information useful to constrain our understanding of galaxy formation and evolution. New telescopes and ever larger and deeper surveys are sure to keep galaxy classifiers busy for many decades to come. Imaging of galaxies with JWST pushes us to higher and higher redshifts, viewing galaxies the early Universe, which reveals the need for new classifications schemes in some cases. The astronomical community's adoption of practice of open data alongside the growth of interest among both students and professionals alike in learning the techniques of ML has resulted in a minor explosion of papers and studies of galaxies morphology using ML algorithms. All of these developments mean this introductory review is sure to become out of date quickly, but hopefully it provides a helpful snapshot of galaxy morphological classification today, alongside links to the terminology used to classify galaxies across the last 100 years. 

\begin{ack}[Acknowledgments]

While this publication does not directly use any of the classifications from  Galaxy Zoo, the author would still like to thank the many thousands of volunteers who have provided classifications at {\tt www.galaxyzoo.org} over the past 15+ years in many phases of the project. Your contributions have inspired me to learn more about galaxy morphology, and to work hard to make sure we are able to maximise the scientific output from the information you provide. 
\end{ack}

\seealso{Ellipticial galaxies and the fundamental plane, The formation of galaxy disks, Mergers of galaxies}

\bibliographystyle{Harvard}
\bibliography{reference}

\end{document}